\documentclass[tex,twocolumn,epjc3,tightenlines]{svjour3}
\RequirePackage[T1]{fontenc}
\RequirePackage{graphicx}
\RequirePackage{mathptmx}      
\RequirePackage[numbers,sort&compress]{natbib}
\RequirePackage[colorlinks,citecolor=blue,urlcolor=blue,linkcolor=blue]{hyperref}
\usepackage{kantlipsum,widetext}
\usepackage{amsmath, amssymb}
\usepackage{graphics,bm}
\usepackage{graphicx}
\usepackage{bbold}
\usepackage{slashed}
\usepackage{feynmf}
\usepackage{wrapfig}
\usepackage{hyperref}

\usepackage{comment} 

\usepackage{tikz}
\usetikzlibrary{arrows,shapes}
\usetikzlibrary{trees}
\usetikzlibrary{matrix,arrows} 			
\usetikzlibrary{positioning}				
\usetikzlibrary{calc,through}				
\usepackage{pgffor}							

\usetikzlibrary{decorations.pathmorphing}	
\usetikzlibrary{decorations.markings}
\tikzset{
    sigmaCT/.style={draw=black, postaction={decorate},
        decoration={markings,mark=at position .99 with {\arrow[draw=black]{>}},mark=at 		 position .99 with {\arrow[draw=black]{<}}}},
    pionCT/.style={dashed,draw=black, postaction={decorate},
        decoration={markings,mark=at position .99 with {\arrow[draw=black]{>}},mark=at position .99 with {\arrow[draw=black]{<}}}},    
    fermionCT/.style={draw=black, postaction={decorate},
        decoration={markings,mark=at position .5 with {\arrow[draw=black]{>}},mark=at position .99 with {\arrow[draw=black]{>}},mark=at position .99 with {\arrow[draw=black]{<}}}},    
    fermion/.style={draw=black, postaction={decorate},
        decoration={markings,mark=at position .25 with {\arrow[draw=black]{>}}}},
    fermionbar/.style={draw=black, postaction={decorate},
        decoration={markings,mark=at position .55 with {\arrow[draw=black]{<}}}},
    pion/.style={dashed,draw=black, postaction={decorate}},
    sigma/.style={draw=black, postaction={decorate}}
}
\newcommand{\be}{\begin{equation}}
\newcommand{\ee}{\end{equation}}

\newcommand{\bqa}{\begin{eqnarray}}
\newcommand{\eqa}{\end{eqnarray}}

\def\sumint{\hbox{$\sum$}\!\!\!\!\!\!\!\!\int}
\def\square{\vcenter{\vbox{\hrule height.4pt
          \hbox{\vrule width.4pt height4pt
          \kern4pt\vrule width.3pt}\hrule height.4pt}}}

\journalname{Eur. Phys. J.}

\begin{document}
\title{
Condensates and pressure of 
  two-flavor chiral perturbation theory at nonzero isospin and temperature}
\author{Prabal Adhikari\thanksref{e1,addr1,addr2} \and Jens O. Andersen\thanksref{e2,addr2} \and Martin A. Mojahed\thanksref{e3,addr2}}
\thankstext{e1}{e-mail: adhika1@stolaf.edu}
\thankstext{e2}{e-mail: andersen@tf.phys.ntnu.no} 
\thankstext{e3}{e-mail: martimoj@stud.ntnu.no}   
\institute{
  St. Olaf College, Faculty of Natural Sciences and Mathematics, Physics Department, 1520 St. Olaf Av.,
  Northfield, MN 55057, United States 
\label{addr1}
          \and
  Department of Physics, 
Norwegian University of Science and Technology, H{\o}gskoleringen 5,
N-7491 Trondheim, Norway
\label{addr2}
}

\date{\today}
\maketitle
\begin{abstract}
We consider two-flavor chiral perturbation theory ($\chi$PT)
at finite isospin chemical potential $\mu_I$ and finite temperature $T$. 
We calculate the effective potential and the quark and pion condensates as functions of $T$ and $\mu_I$
to next-to-leading order in the low-energy expansion in the presence of a pionic source.
We map out the phase diagram in the $\mu_I$--$T$ plane. 
Numerically, we find that the transition to the pion-condensed phase is second order in the region of validity of $\chi$PT, which is in agreement with model calculations and lattice simulations.
Finally, we calculate the pressure to two-loop
order in the symmetric phase for nonzero $\mu_I$ and find that $\chi$PT
seems to be converging very well.
\end{abstract}

\section{Introduction}
Quantum Chromodynamics (QCD) has a very rich phase structure and symmetry-breaking-patterns as a function
of temperature and chemical potentials~\cite{raja,alford,fukurev}.
Normally, the phase diagram is drawn in the $\mu_B$--$T$ plane
and in this case, it includes the hadronic phase, the 
quark-gluon plasma (QGP) phase,
the quarkyonic phase~\cite{rob1}, and various color superconducting phases.

Instead of using a common chemical potential for all quarks, one can
introduce a quark chemical potential $q_f$ for each flavor.
For two flavors, the baryon chemical potential is then
$\mu_B=\frac{3}{2}(\mu_u+\mu_d)$ and the isospin chemical potential
is defined as $\mu_I=\frac{1}{2}(\mu_u-\mu_d)$. 
The phases of QCD now become a function of three control parameters
$T$, $\mu_B$, and $\mu_I$. Restricting ourselves to the
$\mu_I$--$T$ plane, i.e. to $\mu_B=0$ is of particular interest.
In this case, lattice QCD does not suffer from the infamous sign problem
and one can therefore use standard Monte Carlo techniques to
calculate thermodynamic properties and map out the phase diagram
as a function of $T$ and $\mu_I$. This makes it possible to confront
low-energy effective theories such as chiral perturbation
theory~\cite{wein,gasser1,gasser2,bijnens0}
and models such as the Nambu-Jona-Lasinio and
quark-meson models.

\begin{figure}[htb!]
\includegraphics[width=0.45\textwidth]{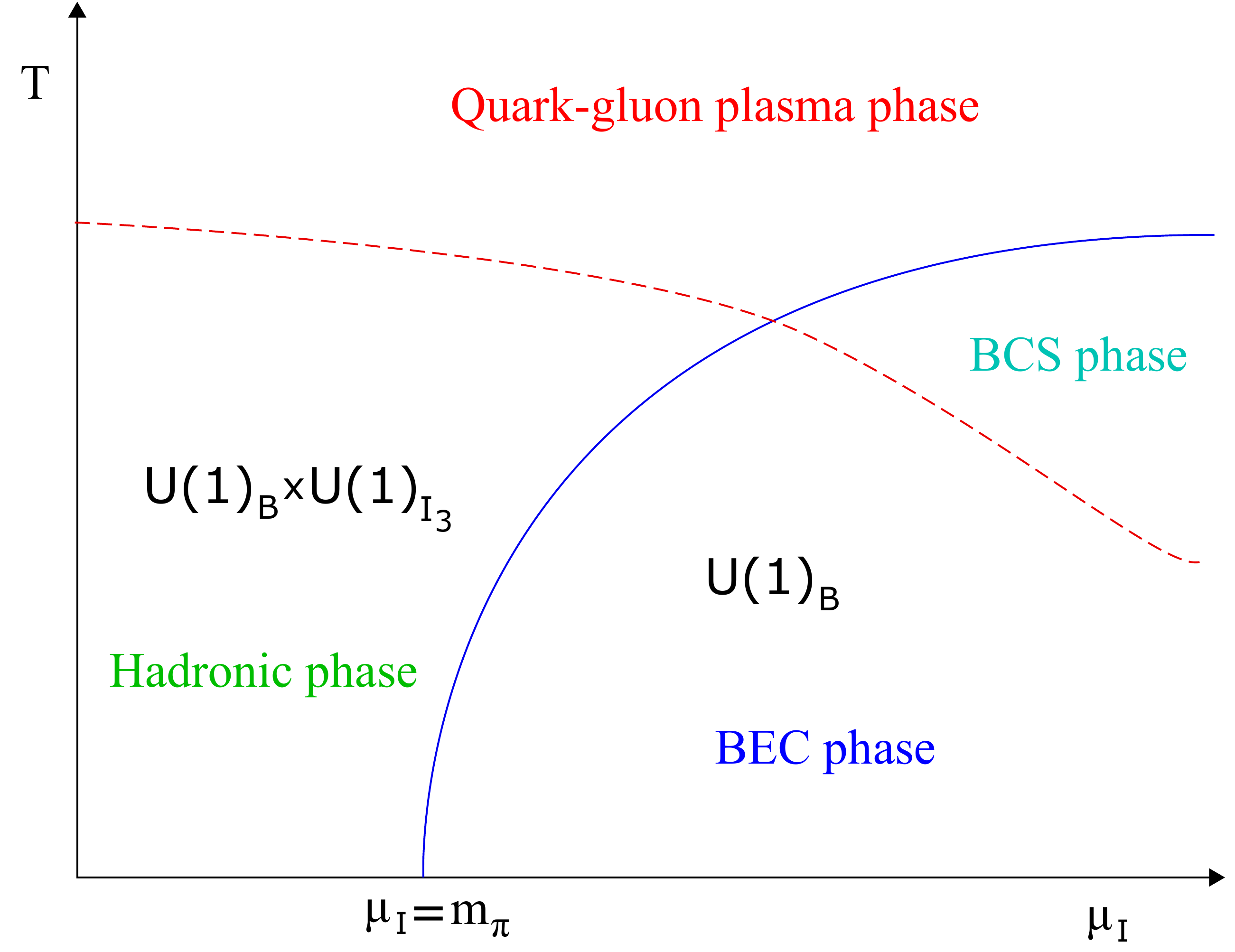}
\caption{Schematic phase diagram in the $\mu_I$--$T$ plane. See main text for details.}
\label{phasediagrams}
\end{figure}

The first simulations of two-flavor QCD
at finite isospin were performed two
decades ago using quenched lattice QCD \cite{kogut1,kogut2}, which was later improved by including dynamical fermions~\cite{kogut3} on relatively coarse lattices.
Later, three-flavor QCD was simulated as well~\cite{kogut4,kogut5} in the phase-quenched approximation.
The nature of the transition from the vacuum phase to a Bose-Einstein condensed
phase of charged pions is second order at $T=0$ and low temperatures.
At larger temperatures, the transition appeared to be first order indicating
the existence of a tricritical point \cite{kogut1,kogut2}.
However, this may be an artifact
of the coarse lattices being used and moreover, the three-flavor
simulations indicated no such point \cite{kogut4,kogut5}.
The recent high-precision lattice
simulations \cite{gergy1,gergy2,gergy3,gergy4} show that the transition is
second order everywhere with critical exponents that are in the $O(2)$
universality class. The phase diagram in the $\mu_I$--$T$ plane is 
sketched in Fig.~\ref{phasediagrams}. 
For small values of the temperature,
and the isospin chemical potential, we are in the confined phase with 
pionic degrees of freedom. The blue line indicates the transition to a 
Bose-condensed phase of charged pions, crossing the $\mu_I$ axis
exactly at $\mu_I=m_{\pi}$.
The red line indicates the transition to a deconfined phase of quarks and gluons.
For small values of $\mu_I$, this is a transition from the confined phase. For larger values of $\mu_I$
it is a transition from the pion-condensed phase to a BCS phase of
weakly interacting quarks that form Cooper pairs~\cite{son}.
The latter is a crossover since it is the same $U(1)$-symmetry
which is broken in the two cases.
Various aspects of the phase diagram at finite $\mu_I$
can be found in e.g. Refs.~\cite{kim,kogut33,loewe,loewe1,njl3f,qmstiele,ueda,lorenz,2fabuki,ebert1,ebert2,toublannjl,bar2f,he2f,heman2,fragaiso,2fbuballa,cohen2,carig0,carigchpt,luca,he3f,ricardo,ruggi}
and a recent review in Ref.~\cite{mannarev}. A first principle dynamical lattice computation of the QCD phase diagram at finite isospin density and temperature for $N_f=4+4$ flavors was reported in Ref. ~\cite{Forcrand}.
In a series of papers, we have studied the properties of
QCD at finite isospin and strange chemical potential at zero temperature
using $\chi$PT~\cite{twoflavor,us,condensates}.
In the case of finite isospin chemical potential and zero strange chemical
potential, the predictions from chiral perturbation theory for a number
of physical quantities have been in good agreement with recent
lattice simulations~\cite{gergy1}. In the present paper, we
generalize our results to finite temperature.
The article is organized as follows.
In Section 2, we discuss the $\chi$PT Lagrangian, the QCD
ground state as a function of isospin chemical potential, and
the fluctuations around it.
In section 3, the effective potential at 
next-to-leading order (NLO)
including a pionic source
as a function of $T$ and $\mu_I$ is calculated.
In section 4, the quark and pion condensates are derived, while
in section 5, we calculate the pressure to next-to-next-to-leading order (NNLO) 
in $\chi$PT in the symmetric
phase. In section 6, we present and discuss our numerical results, including the phase-transition curve separating the normal phase from the pion-condensed phase.
We collect a few useful formulas in an Appendix.

\section{Chiral  Lagrangian, ground state, and fluctuations}
The leading order Lagrangian is
\bqa
\mathcal{L}_{2}=\frac{f^{2}}{4}{\rm Tr}
  \left [\nabla^{\mu} \Sigma^{\dagger} \nabla_{\mu}\Sigma 
    \right ]
+\frac{f^2}{4} {\rm Tr}
\left [\chi^{\dagger}\Sigma+\chi\Sigma^{\dagger}\right ]\; ,
\label{lag0}
\eqa
where $\chi=2B_0M+2iB_0(j_1\tau_1+j_2\tau_2)$ with $M={\rm diag}(m,m)$,
$f$ is the bare pion-decay constant and $2B_0m$ is
the bare pion mass. Moreover, $\tau_a$ are the Pauli matrices and
$j_1$ and $j_2$ are pionic sources which are necessary in order to generate
the pion condensate.
The field $\Sigma$ is written as
\bqa
\Sigma&=&L_{\alpha}\Sigma_{\alpha}R^{\dagger}_{\alpha}\;,
\eqa
where
\bqa
L_{\alpha}&=&A_{\alpha}UA_{\alpha}^{\dagger}\;,\\
R_{\alpha}&=&A_{\alpha}^{\dagger}UA_{\alpha}\;,\\
A_{\alpha}&=&e^{i\frac{\alpha}{2}(\hat{\phi}_1\tau_1+\hat{\phi}_2\tau_2)}\;,\\
U&=&e^{i\frac{\phi_a\tau_a}{2f}}\;,
\eqa
where the subscript $\alpha$ (at tree level)
can be interpreted as a rotation angle of the
quark condensate into a pion condensate, and $\hat{\phi}_i$
are real parameters. The ground state $\Sigma_{\alpha}$
in the pion-condensed phase can be parametrized as~\cite{son}
\bqa
\Sigma_{\alpha}&=&\mathbb{1}\cos\alpha
+i(\hat{\phi}_1\tau_1+\hat{\phi}_2\tau_2)\sin\alpha\;,
\eqa
The parameters satisfy
$\hat{\phi}_1^2+\hat{\phi}_2^2=1$ so that the ground state is properly
normalized, $\Sigma_{\alpha}^{\dagger}\Sigma_{\alpha}=\mathbb{1}$.
Without loss of generality, we can choose 
$\hat{\phi}_1=1$, $\hat{\phi}_2=0$ and $j_1=j$, $j_2=0$.

The covariant derivatives at finite isospin are defined as follows
\bqa
\nabla_{\mu} \Sigma&\equiv&
\partial_{\mu}\Sigma-i\left [v_{\mu},\Sigma \right]\;,\\ 
\nabla_{\mu} \Sigma^{\dagger}&=&
\partial_{\mu}\Sigma^{\dagger}-i [v_{\mu},\Sigma^{\dagger} ]
\;,
\eqa
where $v_{\mu}=\delta_{\mu 0}\mu_{I}\frac{\tau_{3}}{2}$ and $\mu_{I}$ is the
isospin chemical potential.

The Lagrangian is expanded in powers of the field through
quadratic order is
\bqa
{\cal L}_2^{\rm static}&=&f^2(2B_0m_j)
\label{stat}
+\frac{1}{2}f^2\mu_I^2\sin^2\alpha\;,
\\  \nonumber
{\cal L}_2^{\rm linear}&=&
f\left(-2B_0\bar{m}_j
+\mu_I^2\sin\alpha\cos\alpha\right)\phi_1
\\ &&
+f\mu_I\sin\alpha\partial_0\phi_2\;,
\label{lini}
\\ \nonumber
{\cal L}_{2}^{\rm quadratic}&=&\frac{1}{2}\partial_{\mu}\phi_{a}
  \partial^{\mu}\phi_{a}
-  \frac{1}{2}m_a^2\phi_a^2
\\ && 
+\mu_{I}\cos\alpha(\phi_{1}\partial_{0}\phi_{2}-\phi_{2}\partial_{0}\phi_{1})\;.
\label{quadratic}
\eqa
where the source-dependent masses are
\bqa
m_j&=&m\cos\alpha+j\sin\alpha\;,\\
\bar{m}_j&=&m\sin\alpha-j\cos\alpha\;,\\
m_{1}^{2}&=&2B_0m_j -\mu_{I}^{2}\cos{2}\alpha
\;, \\
m_{2}^{2}&=&2B_0m_j -\mu_{I}^{2}\cos^2\alpha
\;, \\
m_{3}^{2}&=&2B_0m_j 
+\mu_{I}^{2}\sin^{2}\alpha\;.
\eqa
The inverse propagator in the $\phi_{a}$ basis is
\bqa
D^{-1}&=
\begin{pmatrix}
D^{-1}_{12}&0\\
0&P^{2}-m_{3}^{2}
\end{pmatrix}\;,\\
D^{-1}_{12}&=
\begin{pmatrix}
P^{2}-m_{1}^{2}&ip_{0}m_{12}\\
-ip_{0}m_{12}&P^{2}-m_{2}^{2}\\
\end{pmatrix}\;,
\eqa
where $m_{12}=2\mu_I\cos\alpha$ and 
$P=(p_0,p)$ is the four-momentum, $P^2=p_0^2-p^2$.
The three dispersion relations are determined by the poles
of the propagator, and the expressions are
\bqa\nonumber
E_{\pi^{\pm}}^{2}&=&p^2+\frac{1}{2}\left(m_{1}^{2}+m_{2}^{2}+m_{12}^{2}\right)
\\
&&\pm\frac{1}{2}\sqrt{4p^{2}m_{12}^{2}+(m_{1}^{2}+m_{2}^{2}
  +m_{12}^{2})^2-4m_{1}^{2}m_{2}^{2}}\;,
\label{pipo}
\\
E_{\pi^0}^{2}&=&p^{2}+m_{3}^{2}\;.
\label{pipo2}
\eqa
At next-to-leading order, the chiral Lagrangian contains a number
of operators~\cite{gasser1}, but not all of them contribute in the present
case. The operators that we need are
\bqa
\nonumber
{\cal L}_4&=&  \frac{1}{4}l_1\left({\rm Tr}
\left[\nabla_{\mu}\Sigma^{\dagger}\nabla^{\mu}\Sigma\right]\right)^2
\\ \nonumber&&
+\frac{1}{4}l_2{\rm Tr}\left[\nabla_{\mu}\Sigma^{\dagger}\nabla_{\nu}\Sigma\right]
    {\rm Tr}\left[\nabla^{\mu}\Sigma^{\dagger}\nabla^{\nu}\Sigma\right]
\\ && \nonumber
+\frac{1}{16}(l_3+l_4)({\rm Tr}[\chi^{\dagger}\Sigma+\Sigma^{\dagger}\chi])^2
\\&& \nonumber
    +\frac{1}{8}l_4{\rm Tr}\left[\nabla_{\mu}\Sigma^{\dagger}\nabla^{\mu}\Sigma\right]
{\rm Tr}[\chi^{\dagger}\Sigma+\Sigma^{\dagger}\chi]
\\ &&
+\frac{1}{2}h_1{\rm Tr}[\chi^{\dagger}\chi] 
\;.
\label{lag}
\eqa
where $l_1$--$l_4$  and $h_1$ are bare coupling constants.
The bare and renormalized couplings $l_i^r(\Lambda)$, $h_i^r(\Lambda)$
are related as follows
\bqa
l_{i}&=&
l_i^r(\Lambda)-\frac{\gamma_i\Lambda^{-2\epsilon}}{2(4\pi)^{2}}
\left[
\frac{1}{\epsilon}+1
\right ]\;,
\label{lowl}
\\
h_i&=&h_i^r(\Lambda)-
\frac{\delta_i\Lambda^{-2\epsilon}}{2(4\pi)^{2}}
\left[\frac{1}{\epsilon}+1\right ]\;,
\label{low2}
\eqa
where $\gamma_i$ and $\delta_i$
are coefficients, and
$\Lambda$ is the renormalization scale in the modified minimal
subtraction ($\overline{\rm MS}$) scheme (see Appendix A).
The renormalized couplings $l_i^r(\Lambda)$ and $h_i^r(\Lambda)$ are running 
satisfying a renormalization group equation.
Since the bare couplings are independent of the renormalization scale
$\Lambda$, differentiation of Eqs.~(\ref{lowl}) --~(\ref{low2})
immediately yields 
\begin{align}
\label{rgrun}
\Lambda{d\over d\Lambda}l_i^r&
=-\frac{\gamma_i\Lambda^{-2\epsilon}}{(4\pi)^2}(1+\epsilon)\;,
&  \Lambda{d\over d\Lambda}h_i^r
=-\frac{\delta_i\Lambda^{-2\epsilon}}{(4\pi)^2}(1+\epsilon)\;.
\end{align}
The low-energy constants $\bar{l}_i$ and $\bar{h}_1$ are defined via
the solutions to the renormalization group equations (\ref{rgrun}) 
for $\epsilon=0$,
  \bqa
  \label{lr}
l_i^r(\Lambda)&=&{\gamma_i\over2(4\pi)^2}\left[\bar{l}_i+\log{M^2\over\Lambda^2}
  \right]\;,\\
h_i^r(\Lambda)&=&{\delta_i\over2(4\pi)^2}\left[\bar{h}_i+\log{M^2\over\Lambda^2}
\right]\;,
\label{hr}
\eqa
and are up to a constant equal to the renormalized couplings
$l_i^r(\Lambda)$  and $h_i^r(\Lambda)$
evaluated at the scale $\Lambda=M$~\cite{gasser1}.
The coefficients $\gamma_i$ and $h_i$ are
\bqa
\gamma_{1}&=&\frac{1}{3}\;,
\hspace{0.3cm}
\gamma_{2}=\frac{2}{3}\;,
\hspace{0.3cm}\gamma_{3}=-\frac{1}{2}\;,
  \\
 \gamma_{4}&=&2\;,  \hspace{0.3cm}\delta_1=0\;.
  \eqa
Since $\delta_1=0$, Eq.~(\ref{hr}) does not apply and 
the coupling $h_1^r$ does not run. 
In the paper~\cite{gasser1}, the authors used another minimal set of
invariant operators than we have listed in Eq.~(\ref{lag}).
The two sets of operators can be transformed into each other using
equations of motion. The couplings are related as 
${h}_1=\tilde{h}_1-\tilde{l}_4$, where the tilde
refers to the original couplings~\cite{scherer}. 
The corresponding values of
the relevant $\tilde{\gamma}_i$ and $\tilde{\delta}_i$ are the same except
$\tilde{\delta}_1=2$. This implies that $\tilde{h}_1^r$ runs according
to Eq.~(\ref{hr}).
At ${\cal O}(p^6)$, there are 53 terms and four contact terms
in the $SU(2)$ chiral Lagrangian~\cite{bijnens0,bijnens1}.
However, only two terms contribute to the static Lagrangian 
${\cal L}_6^{\rm static}$~\cite{hofmann}.
Denoting the bare couplings by $c_i$, only the sum 
$16(c_{10}+2c_{11})(2B_0m)^3$
contributes, where
\bqa\nonumber
c_{10}+2c_{11}&=&
{\Lambda^{-4\epsilon}(c_{10}^{r}+2c_{11}^r)\over f^2}
-{3\Lambda^{-4\epsilon}\over128(4\pi)^4f^2}\left[{1\over\epsilon}+1\right]^2
\\ &&
-{3\Lambda^{-2\epsilon}\over16(4\pi)^2f^2}{l}_3^r\left[{1\over\epsilon}+1\right]
\;,
\label{c10}
\eqa
and where $c_i^r(\Lambda)$ are the renormalized couplings.
Since the bare couplings $c_i$ are independent of the scale $\Lambda$, the
sum $c_{10}^r(\Lambda)+2c_{11}^r(\Lambda)$ satisfies the renormalization 
group equation ($\epsilon=0$)
\bqa
\Lambda{d(c_{10}^r+2c_{11}^r)\over d\Lambda}&=&
-{3\over8}{l_3^r\over(4\pi)^2}
\;.
\eqa
Using Eq.~(\ref{lr}) with $i=3$, we can write
\bqa
\Lambda{d(c_{10}^r+2c_{11}^r)\over d\Lambda}&=&
{3\over32(4\pi)^4}\left[\bar{l}_3+\log{M^2\over\Lambda^2}\right]\;.
\eqa
Integrating this equation and defining the constants $\bar{c}_{10}$
and $\bar{c}_{11}$ as the renormalized couplings $c_{10}^r$ and $c_{11}^r$
at the scale $\Lambda=M$ up to the 
prefactor $-{3\bar{l}_3\over64(4\pi)^4}$~\cite{hofmann}, 
we can write
\bqa\nonumber
c_{10}^r(\Lambda)+2c_{11}^r(\Lambda)
&=&-{3\bar{l}_3\over64(4\pi)^4}(\bar{c}_{10}+2\bar{c}_{11})
-{3\bar{l}_3\over64(4\pi)^4}\log{M^2\over\Lambda^2}
\\ &&
-{3\over128(4\pi)^4}\log^2{M^2\over\Lambda^2}\;.
\label{c10run}
\eqa
The one-loop effective potential in the pion-condensed phase receives a static contribution, which is given by minus the static term in Eq.~(\ref{lag}),
\begin{align}
\label{lagstat1}
\nonumber
\mathcal{L}_{4}^{\rm static}&=(l_{1}+l_{2})\mu_{I}^{4}\sin^4\alpha+l_{4}(2B_0 m_j)\mu_{I}^{2}\sin^{2}\alpha\\  
&+(l_{3}+l_{4})(2B_0m_j)^2
+h_1\left[(2B_0m_j)^2+(2B_0\bar{m}_j)^2\right]\;.
\end{align}
In Sec.~\ref{trykk}, we will calculate the
pressure to order ${\cal O}(p^6)$ in the
symmetric phase, i. e. for $\alpha=0$. In this case the
charged eigenstates and the mass eigenstates coincide and it is
convenient to use the basis
\bqa
\label{complexbasis}
\Phi&=&{1\over\sqrt{2}}(\phi_1+i\phi_2)\;,
\hspace{1cm}
\Phi^{\dagger}={1\over\sqrt{2}}(\phi_1-i\phi_2)\;,
\eqa
We then express the Lagrangian in terms of $\Phi$, $\Phi^{\dagger}$, and
$\phi_3$ instead of $\phi_a$.
The covariant derivatives in the charged basis are defined in the usual way,
\begin{align}
    D_{\mu}\Phi&\equiv (\partial_{\mu}+i\delta_{\mu 0}\mu_I)\Phi,  \\
    D_{\mu}\Phi^{\dagger}&\equiv (\partial_{\mu}-i\delta_{\mu 0}\mu_I)\Phi^{\dagger}\;.
\end{align}
The quadratic Lagrangian Eq.~(\ref{quadratic}) then becomes 
\bqa\nonumber
{\cal L}^{\rm quadratic}_2&=&D_{\mu}\Phi^{\dagger}
D^{\mu}\Phi+
{1\over2}\partial_{\mu}\phi_3\partial^{\mu}\phi_3
\\ &&
-2B_0m\Phi^{\dagger}\Phi-{1\over2}(2B_0m)\phi_3^2\;.
\label{quadro}
\eqa
The quartic terms from Eq.~(\ref{lag0}) can be written as,
\bqa\nonumber
{\cal L}_2^{\rm quartic}&=&
-{1\over3f^2}\Phi^{\dagger}\Phi\left[
D_{\mu}\Phi^{\dagger}D^{\mu}\Phi-B_0m\Phi^{\dagger}\Phi
\right]
\\ && \nonumber
-{1\over3f^2}\phi_3^2\left[
D_{\mu}\Phi^{\dagger}D^{\mu}\Phi-2B_0m\Phi^{\dagger}\Phi
\right]
\\ && \nonumber
-{1\over6f^2}\Phi^{\dagger}\Phi\left[2\partial_{\mu}\phi_3\partial^{\mu}\phi_3
+2B_0m\phi_3^2\right]
\\ \nonumber
&&+{1\over6f^2}\left[\Phi\Phi D_{\mu}\Phi^{\dagger}D^{\mu}\Phi^{\dagger}
+\Phi^{\dagger}\Phi^{\dagger}D_{\mu}\Phi D^{\mu}\Phi\right]
\\
&&+{1\over6f^2}\left(\partial_{\mu}\phi_3^2\right)\partial^{\mu}\left(\Phi^{\dagger}\Phi\right)
+{2B_0m\over24f^2}\phi_3^4\;.
\label{quartic1}
\eqa
The quadratic terms from Eq.~(\ref{lag}) are
\bqa
\nonumber
{\cal L}_4^{\rm quadratic}&=&
{2l_4\over f^2}\left[D_{\mu}\Phi^{\dagger}
D^{\mu}\Phi+\frac{1}{2}\partial_{\mu}\phi_3\partial^{\mu}\phi_3\right]2B_0m
\\ &&
-{(l_3+l_4)\over f^2}\left[2\Phi^{\dagger}\Phi+\phi_3^2\right](2B_0m)^2\;.
\label{termcounter}
\eqa

\section{Effective potential with pionic source}
In this section, we calculate the effective potential to NLO
including a pionic source. At $T=0$, this calculation was carried out
in Ref.~\cite{condensates}. It is straightforward to generalize the
result to finite temperature and we include the calculation for completeness.
We perform the finite-temperature calculations using the
imaginary-time formalism. The energy $\omega$ is then replaced
by $iP_0$, where the (bosonic) Matsubara frequencies are given by
$P_0=2\pi nT$, $n\in \mathbb{Z}$. The Minkowski-space propagator ${i\over P^2-m^2}$
is replaced by a Euclidean propagator ${1\over P^2+m^2}$, where
$P^2=P_0^2+p^2$.
The propagator for the complex field in the symmetric phase is
$\Delta={1\over(P_0+i\mu_I)^2+p^2+M^2}$.
The source-dependent one-loop contribution to the
effective potential can be written as
\bqa\nonumber
V_{1}
&=&{1\over2}\sumint_P\log\left[P_0^2+E_{\pi^+}^2\right]
+{1\over2}\sumint_P\log\left[P_0^2+E_{\pi^-}^2\right]\nonumber
\\ && \nonumber
+{1\over2}\sumint_P\log\left[P_0^2+E_{\pi^0}^2\right]
\\ \nonumber
&=&{1\over2}\int_p\left[E_{\pi^+}+E_{\pi^-}+E_{\pi^0}\right]
+T\int_p\log\left[1-e^{-\beta E_{\pi^{+}}}\right]
\\ &&
+T\int_p\log\left[1-e^{-\beta E_{\pi^{-}}}\right]
+T\int_p\log\left[1-e^{-\beta E_{\pi^0}}\right]\;,
\eqa
where the dispersion relations $E_{\pi^0}$ and $E_{\pi^{\pm}}$
are given by Eqs.~(\ref{pipo})--(\ref{pipo2}). 

The zero-temperature integrals involving the charged excitations  cannot
be done analytically in dimensional regularization. However, we can
isolate the ultraviolet divergences by adding and subtracting
appropriate terms that can be calculated in dimensional regularization.
The ultraviolet behavior of $E_{\pi^{\pm}}$ is given by
$E_{1,2}=\sqrt{p^2+m^2_{1,2}+{1\over4}m_{12}^2}$, i.e.
excitations with masses
$\tilde{m}_1^2=2B_0m_j+\mu_I^2\sin^2\alpha=m_3^2$
and $\tilde{m}_2^2=2B_0m_j$. We can therefore write
\bqa
V_{1}&=&V_1^{\rm div}+V_{1}^{\rm fin}\;,
\eqa
where
\bqa\nonumber
V_1^{\rm div}&=&\int_p\sqrt{p^2+m_3^2}+{1\over2}\int_p\sqrt{p^2+\tilde{m}_2^2},
\\ \nonumber
&=&-{1\over2(4\pi)^2}\left[
{1\over\epsilon}+{3\over2}+\log{2B_0m\over{m}_3^2}
\right](2B_0m_j+\mu_I^2\sin^2\alpha)^2
\\ &&
-{1\over4(4\pi)^2}\left[
{1\over\epsilon}+{3\over2}+\log{2B_0m\over{\tilde{m}_2^2}}
\right](2B_0m_j)^2\;,
\label{v11}
\\ 
V_1^{\rm fin}&=&{1\over2}\int_p\left[E_{\pi^+}+E_{\pi^-}
  -\sqrt{p^2+\tilde{m}_1^2}-\sqrt{p^2+\tilde{m}_2^2}
\right]
\;.
\label{v12}
\eqa
The complete NLO effective potential is the sum of Eqs. (\ref{v11})--(\ref{v12}) minus Eqs.~(\ref{stat}) and (\ref{lagstat1}).
Renormalization is carried out by replacing $l_i$ by $l_r(\Lambda)$
according to Eqs.~(\ref{lowl}) and (\ref{lr}).
\begin{widetext}
\noindent
This yields
\bqa\nonumber
V_{\rm eff}&=&-f^2(2B_0m_j)-{1\over2}f^2\mu_I^2\sin^2\alpha \\
\nonumber
&&-\frac{1}{4(4\pi)^{2}}\left[\frac{3}{2}-\bar{l}_{3}+4\bar{l}_{4}
  +\log\left(\frac{M^2}{\tilde{m}_{2}^{2}}\right)
    +2\log\left(\frac{M^2}{m_{3}^{2}}\right)\right ](2B_{0}m_{j})^{2}
-\frac{1}{2(4\pi)^{2}}\left[1+2\bar{l}_{4}
  +2\log\left(\frac{M^2}{m_{3}^{2}}\right) \right]
(2B_{0}m_{j})\mu_{I}^{2}\sin^{2}\alpha\\ \nonumber
&&-\frac{1}{2(4\pi)^{2}}\left[\frac{1}{2}+\frac{1}{3}\bar{l}_{1}+\frac{2}{3}
  \bar{l}_{2}+\log\left(\frac{M^2}{m_{3}^{2}}\right) \right ]\mu_{I}^{4}
\sin^{4}\alpha
-{1\over(4\pi)^2}\bar{h}_1\left[(2B_0m_j)^2+(2B_0\bar{m}_j)^2\right]
+V^{\rm fin}_{1,\pi^{+}}+V^{\rm fin}_{1,\pi^{-}}\\
&&+T\int_p\log\left[1-e^{-\beta E_{\pi^{+}}}\right]
+T\int_p\log\left[1-e^{-\beta E_{\pi^{-}}}\right]
+T\int_p\log\left[1-e^{-\beta E_{\pi^0}}\right]
\; ,
\label{renpi}
\eqa
\end{widetext}
where we have defined
$\bar{h}_{1}=(4\pi)^{2}h_{1}$. We note that the result is independent of $\Lambda$, which implies that thermodynamic functions and condensates that we generate are independent of the renormalization scale.
\section{Quark and pion condensates}
In Ref.~\cite{condensates}, we derived the quark and pion condensates
in the pion-condensed phase at $T=0$. In this section, we generalize
the results to finite temperature. We begin with the definition of the chiral condensate and the pion condensate, which are
\bqa
\langle\bar{\psi}\psi\rangle_{\mu_I}
&=&{1\over2}{\partial V_{\rm eff}\over\partial m}\;,
\hspace{0.4cm}
\langle\pi^{+}\rangle_{\mu_I}={1\over2}{\partial V_{\rm eff}\over\partial j}\;,
\eqa
respectively, where the subscript is a reminder that the condensates depend on the
isospin chemical potential. At leading order the condensates read
\bqa
\langle\bar{\psi}\psi\rangle_{\mu_I,\rm tree}&=-f^2B_0\cos\alpha\;,\\
\langle\pi^{+}\rangle_{\mu_I,\rm tree}&=-f^2B_0\sin\alpha\;.
\eqa
with the pion condensate vanishing in the normal vacuum with $\alpha=0$. The temperature dependence of the effective potential enters through loop corrections, so all tree-level results are temperature independent. Furthermore, we also note that the sum of the square of the condensates is constant,
\begin{equation}
\langle\bar{\psi}\psi\rangle_{\mu_I,\rm tree}^2+\langle\pi^{+}\rangle_{\mu_I,\rm tree}^2=(-f^2B_0)^2\;,
\label{rotatie}
\end{equation}
with a radius equal to the chiral condensate of the normal vacuum. At NLO, this rotation relation is violated and the condensates
can be expressed as
\begin{align}
\langle\bar{\psi}\psi\rangle_{\mu_I}&=\langle\bar{\psi}\psi\rangle_{\mu_I,0}+\langle\bar{\psi}\psi\rangle_{\mu_I,T}\;, \\
\langle\pi^{+}\rangle_{\mu_I}&=\langle\pi^{+}\rangle_{\mu_I,0}+
\langle\pi^{+}\rangle_{\mu_I,T}\;,
\end{align}
where the first terms are the temperature-independent contributions while the second terms are temperature dependent. These contributions are calculated using the effective potential in Eq.~(\ref{renpi}). The results below are obtained by first taking the appropriate partial derivatives and then setting the reference scale $M^2$ equal to $2B_0m$.

\begin{widetext}
\bqa\nonumber
\langle\bar{\psi}\psi\rangle_{\mu_{I},0}&=&
-{f}^2{B}_0
\cos\alpha\left[1+  {1\over2(4\pi)^2}\left(-\bar{l}_3+4\bar{l}_4
   +  \log{2B_0m\over\tilde{m}_2^2}
   +2\log{2B_0m\over m_3^2}       \right){2B_0m_j\over f^2}
  +{1\over(4\pi)^2}\left(\bar{l}_4+\log{2B_0m\over m_3^2}\right)
  {\mu_I^2\sin^2\alpha\over f^2}
\right]
\\ && 
-{2\over(4\pi)^2}\bar{h}_1{B}_0(2B_0m)
+{1\over2}{\partial V_{\rm 1,\pi^+}^{\rm fin}\over\partial m}
+{1\over2}{\partial V_{\rm 1,\pi^-}^{\rm fin}\over\partial m}\;.
\label{light} \\ \nonumber
\langle\pi^+\rangle_{\mu_{I},0}&=&
-{f}^2{B}_0
\sin\alpha\left[1+  {1\over2(4\pi)^2}
  \left(-\bar{l}_3+4\bar{l}_4
    +\log{2B_0m\over\tilde{m}_2^2}
    +2\log{2B_0m\over m_3^2} 
  \right){2B_0m_j\over f^2}
  +{1\over(4\pi)^2}\left(\bar{l}_4+\log{2B_0m\over m_3^2}\right)
    {\mu_I^2\sin^2\alpha\over f^2}
  \right]
\\  && 
{-{4\over(4\pi)^2}\bar{h}_1{B}_0^2{j}}
+
{1\over2}
{\partial V_{\rm 1,\pi^+}^{\rm fin}\over\partial j}
+{1\over2}{\partial V_{\rm 1,\pi^-}^{\rm fin}\over\partial j}\;,\\
\label{pioncon}
\langle\Bar{\psi}\psi\rangle_{\mu_I,T}&=&
    \frac{B_0\cos\alpha}{2}\left\{\int_k\frac{n_B\left(E_{\pi^+}\right)}
    {E_{\pi^+}}
    \left[1+\frac{m_{12}^2}{\sqrt{4k^2m_{12}^2+(m_1^2+m_2^2+m_{12}^2)^2-4m_1^2m_2^2}}
    \right]\nonumber \right. \\
  &&\left.+\int_k\frac{n_B\left(E_{\pi^-}\right)}{E_{\pi^-}}\left[1-\frac{m_{12}^2}
    {\sqrt{4k^2m_{12}^2+(m_1^2+m_2^2+m_{12}^2)^2-4m_1^2m_2^2}}\right]+\int_k
    \frac{n_B\left(E_{\pi^0}\right)}{E_{\pi^0}} \right\}\;,\\
  \langle\pi^+\rangle_{\mu_I,T}&=& 
  \frac{B_0\sin\alpha}{2}\left\{\int_k\frac{n_B\left(E_{\pi^+}\right)}
    {E_{\pi^+}}\left[1+\frac{m_{12}^2}{\sqrt{4k^2m_{12}^2+(m_1^2+m_2^2+m_{12}^2)^2
    -4m_1^2m_2^2}}\right]\nonumber \right. \\
&  &\left.+\int_k\frac{n_B\left(E_{\pi^-}\right)}{E_{\pi^-}}\left[1-\frac{m_{12}^2}
    {\sqrt{4k^2m_{12}^2+(m_1^2+m_2^2+m_{12}^2)^2-4m_1^2m_2^2}}\right]+\int_k
    \frac{n_B\left(E_{\pi^0}\right)}{E_{\pi^0}} \right\}\;.
    \label{pionconfiniteT}
\eqa
\end{widetext}
The temperature-independent contributions
were first calculated in Ref.~\cite{condensates}, while
the temperature-dependent contributions are new
and follow directly from the last line of
Eq.~(\ref{renpi}).

\section{Two-loop pressure in the symmetric phase}
\label{trykk}
In this section, we calculate the pressure to two-loop order
which corresponds to a next-to-next-to-leading order calculation in the
low-energy expansion. 
Through one-loop, the pressure is given by
\bqa\nonumber
{\cal P}_{0+1}&=&
f^2(2B_0m)
+(l_3+l_4)(2B_0m)^2+h_1(2B_0m)^2
\\ && \nonumber
-{1\over2}\sumint_Q\log[Q^2+2B_0m]
\\  &&
-\sumint_Q\log[(Q_0+i\mu_I)^2+q^2+2B_0m]
\;,
\label{1l}
\eqa
where the mean-field term Eq.~(\ref{stat}) and the contribution from the
counterterm Eq.~(\ref{lagstat1}) are evaluated at $\alpha=0$.
Here and in the following, the subscript $n$ 
of ${\cal P}_n$ denotes the $n$th order in the low-energy expansion and
the subscript $0+1..+n$ denotes the complete result to the same order.
Using the expressions for the sum-integral (\ref{sumint22})
as well renormalizing the couplings using Eqs.~(\ref{lowl})--(\ref{low2}), and
the relation Eq.~(\ref{lr}), we obtain the pressure to NLO 
  \bqa\nonumber
  {\cal P}_{0+1}&=&f^2(2B_0m)
+{(2B_0m)^2\over4(4\pi)^2}\left[\frac{3}{2}-\bar{l}_3+4\bar{l}_4
+4\bar{h}_1
\right]\\ && \nonumber
-T\int_q\left\{\log\left[1-e^{-\beta E_q}\right]
+    \log\left[1-e^{-\beta(E_q-\mu_I)}\right]
\right.\\&&
\left.
+    \log\left[1-e^{-\beta(E_q+\mu_I)}\right]
  \right\}\;.
\label{1loop}
  \eqa
The first line in Eq.~(\ref{1loop}) is the vacuum energy, while the second and
third line are the pressure of an ideal gas of massive particles of mass $2B_0m$
at finite isospin chemical potential.

\begin{figure}[htb]
\begin{center}
\begin{tikzpicture}[line width=1.2 pt, scale=1]
	
  \draw[fermion] (10.5,0.5) arc (360:0:.5);
  	\draw[fermion] (11.5,0.5) arc (360:0:.5);
	
	\draw[fermion] (13.5,0.5) arc (360:0:.5);
	\draw[pion] (14.5,0.5) arc (360:0:.5);                

	\draw[pion] (16.5,0.5) arc (360:0:.5);
	\draw[pion] (17.5,0.5) arc (360:0:.5);

        \draw[fermion] (10.5,-1.) arc (360:0:.5);
        \filldraw[black] (10.5,-1.) circle (2pt);

	\draw[pion] (13.5,-1) arc (360:0:.5);                
        \filldraw[black] (13.5,-1.) circle (2pt);        
        \filldraw[black] (15.5,-1.) circle (2pt);        
      \end{tikzpicture}
      \caption{NNLO diagrams for the pressure in the symmetric phase.
  A solid line corresponds to the charged pion and a
  dashed line corresponds to the neutral pion. A black dot
is a mass insertion or a temperature-independent counterterm.}
\label{dia2}
\end{center}
\end{figure}
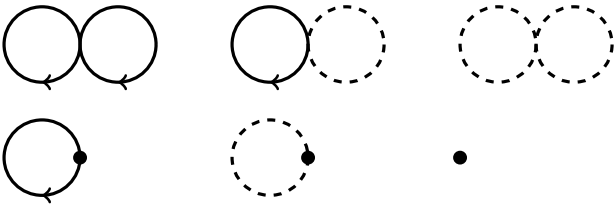

At NNLO, there are a number of diagrams that contribute to the pressure.
The two-loop diagrams arising from ${\cal L}_2^{\rm quartic}$, Eq.~(\ref{quartic1})
are shown in the first line of Fig.~(\ref{dia2}), while the
the counterterm diagrams arising from Eq.~(\ref{termcounter})
as well as the contact term are shown in the second line of Fig.~(\ref{dia2}).
Using Eqs.~(\ref{sumint22})--(\ref{sumint33}), 
the two-loop contribution to the pressure can be written as
\bqa\nonumber
{\cal P}_2^{\rm quartic}&=&-{2B_0m\over2f^2}\sumint_K{1\over K^2+2B_0m}
\sumint_Q{1\over(Q_0+i\mu_I)^2+q^2+2B_0m}
\\ && \nonumber
+{2B_0m\over8f^2}\left[\sumint_Q{1\over Q^2+2B_0m}\right]^2
\\ && \nonumber
+{1\over f^2}\left[\sumint_Q{(Q_0+i\mu_I)\over(Q_0+i\mu_I)^2+q^2+2B_0m}\right]^2
\\ \nonumber
&=&-{3(2B_0m)^3\over8(4\pi)^4f^2}\left[{1\over\epsilon^2}+{2\over\epsilon}\left(1+\log{\Lambda^2\over 2B_0m}\right)
\right.\\ &&\left.\nonumber
+3+{\pi^2\over6}
+4\log{\Lambda^2\over 2B_0m}
+2\log^2{\Lambda^2\over 2B_0m}
\right] \nonumber
\\ \nonumber&&
+{(2B_0m)^2\over4(4\pi)^2f^2}
\left({1\over\epsilon}+1+\log{\Lambda^2\over 2B_0m}\right)
\\ \nonumber &&\times
\int_q\left[{n_B(E_q)\over E_q}+{n_B(E_q-\mu_I)\over E_q}
+{n_B(E_q+\mu_I)\over E_q}\right]
\\ \nonumber
&&-{2B_0m\over4f^2}
\int_q\left[{n_B(E_q-\mu_I)\over E_q}
+{n_B(E_q+\mu_I)\over E_q}
\right]
\\ && \nonumber
\times \int_q{n_B(E_k)\over E_k}
+{2B_0m\over8f^2}\left[\int_q{n_B(E_q)\over E_q}\right]^2
\\ &&
-{1\over4 f^2}\left\{\int_q\big[n_B(E_q-\mu_I)-n_B(E_q+\mu_I)
\big]\right\}^2\;.
\label{2l}
\eqa
The one-loop 
counterterm
contribution to the pressure ${\cal P}_2^{\rm ct}$
from ${\cal L}_4^{\rm quadratic}$ is
\bqa\nonumber
{\cal P}_2^{\rm ct}&=&-{l_3(2B_0m)^2\over f^2}
\sumint_Q\left[
{2\over(Q_0+i\mu_I)^2+q^2+2B_0m}
+{1\over Q^2+2B_0m}
\right]\\ \nonumber
&=&\frac{3l_3(2B_0m)^3}{(4\pi)^2f^2}
\left[{1\over\epsilon}+1+\log{\Lambda^2\over 2B_0m}
+{\pi^2+12\over12}\epsilon
+\log{\Lambda^2\over2B_0m}\epsilon
\right. \\ &&\left. \nonumber
+\frac{1}{2}\log^2\left(\frac{\Lambda^2}{2B_0m}\right)\epsilon
\right]
-{l_3(2B_0m)^2\over f^2}
\\ &&\times
\int_q\left[{n_B(E_q)\over E_q}+{n_B(E_q-\mu_I)\over E_q}
+{n_B(E_q+\mu_I)\over E_q}\right]\;,
\label{IC}
\eqa
where we have expanded the zero-temperature part of the loop integral to
order $\epsilon$ in order to pick up a finite term when it
is multiplied by the bare coupling $l_3$.
Finally, the contact term is
\bqa
{\cal P}_{2}^{\rm contact}&=&
16(c_{10}+2c_{11})(2B_0m)^3\;.
\label{contact}
\eqa
The NNLO contribution to the pressure is then given by the sum of
Eqs.~(\ref{2l}), (\ref{IC}), and (\ref{contact}).
The ultraviolet divergences are eliminated 
upon substituting $l_3$ by $l_3^r$
using (\ref{lr}) and $c_{10}+2c_{11}$ by $c_{10}^r+2c_{11}^r$ using
Eq.~(\ref{c10}). The running couplings 
$l_3^r$ and  
$c_{10}^r+2c_{11}^r$ are then replaced by the right-hand side of Eqs.~(\ref{lr}) 
and~(\ref{c10run}).
After adding Eqs.~(\ref{1loop})--(\ref{contact}) and renormalizing, we obtain
\bqa\nonumber
{\cal P}_{0+1+2}&=&
f^2(2B_0m)
+{(2B_0m)^2\over4(4\pi)^2}\left[{3\over2}-\bar{l}_3+4\bar{l}_4+4\bar{h}_1
\right]
\\&& \nonumber
+{3\bar{l}_3(2B_0m)^3\over4(4\pi)^4f^2}\left[\bar{c}_{10}+2\bar{c}_{11}\right]
\\ && \nonumber
-T\int_q\left\{\log\left[1-e^{-\beta E_q}\right]
+    \log\left[1-e^{-\beta(E_q-\mu_I)}\right]
\right. \\ &&\left. \nonumber
+    \log\left[1-e^{-\beta(E_q+\mu_I)}\right]
\right\}
\\ && \nonumber
{+}{(2B_0m)^2\over4(4\pi)^2f^2}\bar{l}_3\\ && \nonumber
\times
\int_q\left[{n_B(E_q)\over E_q}+{n_B(E_q-\mu_I)\over E_q}
+{n_B(E_q+\mu_I)\over E_q}\right]
\\ && \nonumber
-{2B_0m\over4f^2}\left[\int_q{n_B(E_q-\mu_I)\over E_q}+{n_B(E_q+\mu_I)\over E_q}
\right]
\\ && \nonumber
\times\int_k{n_B(E_k)\over E_k}
+{2B_0m\over8f^2}\left[\int_q{n_B(E_q)\over E_q}\right]^2
\\ &&
-{1\over4f^2}\left\{\int_q\big[n_B(E_q-\mu_I)-n_B(E_q+\mu_I)\big]\right\}^2\;.
\label{pressure11}
\eqa
The term on the third and fourth line can be absorbed in the one-loop result (\ref{1loop})
by making the substitution $2B_0m\rightarrow m^2_{\pi}=2B_0m\left(1-{2B_0m\over2(4\pi)^2f^2}\bar{l}_3\right)$, 
where $m_{\pi}$
is the physical 
pion mass in the vacuum, at one loop.
This can be seen by writing $m_{\pi}^2=m^2+\delta m^2$ and 
expanding the one-loop result to first order in $\delta m^2$.
The final result then reads
\bqa\nonumber
{\cal P}_{0+1+2}&=&
f^2(2B_0m)+{(2B_0m)^2\over4(4\pi)^2}\left[{3\over2}-\bar{l}_3+4\bar{l}_4+4\bar{h}_1
\right]
\\ \nonumber
&&+{3\bar{l}_3(2B_0m)^3\over4(4\pi)^4f^2}\left[\bar{c}_{10}+2\bar{c}_{11}\right]
\\ && \nonumber
-T\int_q\left\{\log\left[1-e^{-\beta E_q}\right]
+    \log\left[1-e^{-\beta(E_q-\mu_I)}\right]
\right. \\ &&\left. \nonumber
+    \log\left[1-e^{-\beta(E_q+\mu_I)}\right]
\right\}
\\ && \nonumber
-{2B_0m\over4f^2}
\int_q\left[{n_B(E_q-\mu_I)\over E_q}+{n_B(E_q+\mu_I)\over E_q}\right]
\\ && \nonumber
\times\int_p{n_B(E_p)\over E_p}
+{2B_0m\over8f^2}\left[\int_q{n_B(E_q)\over E_q}\right]^2
\\ && 
-{1\over4f^2}\left\{\int_q\big[n_B(E_q-\mu_I)-n_B(E_q+\mu_I)\big]\right\}^2\;.
\label{pressure121}
\eqa
The final result is scale independent.
In the limit $\mu_I\rightarrow0$, the temperature-dependent terms of the
result Eq.~(\ref{pressure121}) reduce to
the result of Gerber and Leutwyler when restricting their three-loop result to
two loops \cite{gerber}. 
In the chiral limit, it reduces to a gas of noninteracting
bosons.~\footnote{In the chiral limit, $\mu_I=0$ in order to remain
in the symmetric phase. The first correction to the ideal-gas result
is of order $T^8/f^4$~\cite{gerber}.}
The temperature-independent terms agree with the result first obtained
in Ref.~\cite{hofmann} for the full vacuum energy to order ${\cal O}(p^6)$
in $\chi$PT.

\section{Numerical results and discussion}
Our results for the quark and pion condensates, 
and the pressure contain a number  parameters from the chiral Lagrangian,
namely $2B_0m$ and $f$, the four low-energy constants $l_1-l_4$, and
the contact parameter $h_1$.
The low-energy constants $\bar{l}_1$ and $\bar{l}_2$ were measured
experimentally via $d$-wave scattering lengths, while
$\bar{l}_3$ has been estimated using three-flavor
QCD \cite{gasser1}. The low-energy constant $\bar{l}_4$ is related to
the scalar radius of the pion. 
We have determined $\bar{h}_1$ using the values and 
uncertainties of the three-flavor low-energy constants in 
Ref.~\cite{Hr2ref1} and 
the mapping of three-flavor LECs to two-flavor LECs 
as discussed in Ref.~\cite{gasser2}.~\footnote{There  is  another  choice  of $\bar{h}_1$ in 
the literature~\cite{gasser1}, 
which happens to be model-dependent (with calculations based on $\rho$-dominance).}
The numerical values are~\cite{cola}
\begin{align}
  \label{variasjon1}
\bar{l}_{1}&=-0.4\pm 0.6\;,
&\bar{l}_{2}=4.3\pm 0.1\;,\\
\label{variasjon3}
\bar{l}_{3}&=2.9 \pm 2.4\;,
&\bar{l}_{4}=4.4 \pm 0.2\;,\\
\bar{h}_1&=-1.5\pm0.2.
\label{variasjon2}
\end{align}
We are aware of more recent values of the low-energy 
constants~\cite{newlow},
but for consistency, we use the same values as 
Refs.~\cite{twoflavor,us,condensates} to compare $\chi$PT with LQCD at zero temperature.
The relations between the bare parameters $f$ and $2B_0m$ in the
chiral Lagrangian and the physical pion mass $m_{\pi}$
and the pion-decay constant $f_{\pi}$ at one loop
are given by~\cite{gasser1}
\bqa
m_{\pi}^2&=&2B_0m\left[1-{2B_0m\over2(4\pi)^2f^2}\bar{l}_3\right]\;,
\label{mpi}
\\ 
f_{\pi}^2
&=&f^2\left[1+{4B_0m\over(4\pi)^2f^2}\bar{l}_4\right]\;.
\label{fpi}
\eqa
Thus once we know the couplings $\bar{l}_3$ and $\bar{l}_4$ as well as
the pion mass and the pion-decay constant, we can determine the parameters
$f$ and $2B_0m$ in the chiral Lagrangian. As in the previous
papers~\cite{twoflavor,us,condensates}, we adopt the values of
$m_{\pi}$ and $f_{\pi}$ used in the lattice simulations of
Refs.~\cite{gergy1,gergy2,gergy3},
\bqa
m_{\pi}&=&131\pm3 {\rm\ MeV}\;,
\hspace{1cm}
f_{\pi}={128\pm3\over\sqrt{2}}{\rm\ MeV}\;.
\eqa
In the remainder of the paper, we shall be using only the central values
of the parameters. 
Using the central values of the values quoted above, we obtain
the bare values 
\bqa
\label{pp1}
  (2B_0m)_{\rm cen}&=&132.4884\,{\rm\ MeV}\;,
\hspace{0.4cm}
  f_{\rm cen}=84.9342\,{\rm\ MeV}\;.
\label{pp3}
\eqa
Using the value (\ref{pp1}), we can calculate $B_0$ using the continuum value of the quark mass, the bare pion-decay constant and the bare pion mass. Unfortunately, the quark mass is not known for the
isospin simulations of Refs.~\cite{gergy1,gergy2,gergy3}.
We therefore use the continuum value of
the quark mass from a different lattice computation~\cite{BMW}, similar to what we did in the zero temperature analysis of Ref~\cite{condensates}. Previously, the zero temperature condensates~\cite{condensates} were found to be insensitive to the values of the continuum quark mass (and their corresponding uncertainties)~\cite{BMW}. The parameters most sensitive to uncertainties were found to be the pion mass and pion-decay constant.

In order to study the phase transition (critical isospin chemical 
potential
as a function of temperature) analytically, we
perform a Ginzburg-Landau expansion of the effective potential in powers of $\alpha$ (around $\alpha=0$),
\bqa
V_{\rm GL}&=&V_{\rm eff}(0)+a_2\alpha^2+a_4\alpha^4+...
\eqa
At $T=0$, the coefficients $a_i$ are functions of $\mu_I$ and the
parameters of the Lagrangian. At finite temperature, there is additional
temperature dependence. The critical isospin chemical potential
$\mu_I^c$ as a function of the critical temperature $T^c$ is defined as the curve in the phase diagram where $a_2$ vanishes.
If $a_4$ evaluated at a point on the phase-transition curve is larger than zero, then the
transition is second order at that point. If $a_4$ is smaller than zero, then the
transition is first order at that point.
Finally, if $a_4=0$, it is a tricritical point where the order of the phase transition changes its character from first to second order.
In Ref.~\cite{twoflavor}, it was shown that
$a_2={1\over2}f_{\pi}^2[m_{\pi}^{2}-\mu_I^2]$ at $T=0$, which yields a critical chemical
potential of $\mu_I^c=m_{\pi}$. At this value of $\mu_I$, $a_4$ is positive and consequently the transition is second order.

A low-temperature expansion of the effective potential
was carried out in Ref.~\cite{split2} in the context of two-color QCD to find the coefficients $a_2$ and $a_4$. In the same paper, the authors also applied these techniques to three-color QCD at finite isospin chemical potential within the approximation
$(\mu_I-m_{\pi})\ll T\ll m_{\pi}$. Using $m_{\pi}=131{\rm\ MeV}$, the approximation holds within the $20\%$ level for $T\le 26.2{\rm\ MeV}$ and $\frac{\mu_{I}-m_{\pi}}{m_{\pi}}\le 0.04$, with the bounds halved assuming a $10\%$ threshold.
They obtained the following critical isospin chemical potential, where $a_2=0$, for two-flavor QCD
\bqa
\mu_I^c(T)&=&m_{\pi}+\frac{1}{4f_{\pi}^2}\sqrt{\frac{m_{\pi}^3T^3}{2\pi^3}}
\zeta\left(\mbox{$3\over2$}\right).
\label{ana}
\eqa
The authors also obtained a tricritical point, where $a_2=a_4=0$, at the following temperature
\bqa
T_{\rm tri}=2m_{\pi}\frac{4\pi-\zeta(\frac{1}{2})\zeta(\frac{3}{2})}{3\zeta^2(\frac{3}{2})}\approx 1.6m_{\pi}\;.
\eqa
However, $T_{\rm tri}$ is outside the region of validity of the approximations
used to obtain the result.
\begin{figure}[htb]
\includegraphics[width=0.45\textwidth]{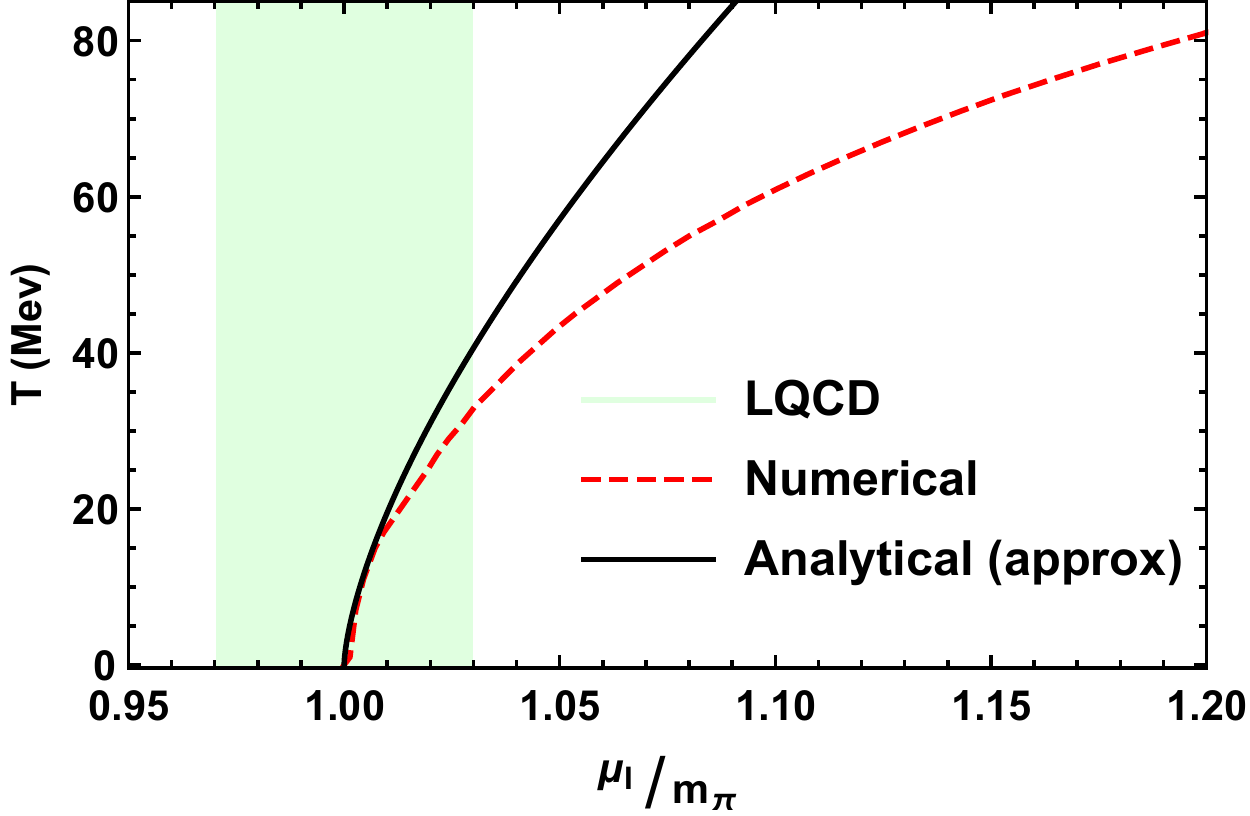} 
\caption{Phase diagram of chiral perturbation theory at finite isospin chemical potential and temperature using the central values of parameters.
See main text for details.}
\label{phasediagram}
\end{figure}

A first-order transition is signalled by a jump in the order-parameter $\langle{\pi^+}\rangle$. From Eq. (\ref{pionconfiniteT}), it is easily seen that a jump in $\langle{\pi^+}\rangle$ can only be generated by a jump in $\alpha$, as long as $\mu_I$ and $T$ are continuous variables. Numerically, we do not find any signs of such a jump for temperatures up to
$100$ MeV, which is well above the temperature range where
the $\chi$PT critical isospin chemical potential agrees with that from lattice QCD.
Finally, lattice results~\cite{gergy3,gergy4} as well as model
calculations see e.g.~\cite{heman2,njl3f,allofus} find a second-order transition everywhere.

The phase boundary itself is shown in Fig.~\ref{phasediagram}. The
red dashed line is the numerical result for the onset of pion condensation as a function
of temperature, the solid black line is the analytical result of Eq. (\ref{ana}), and the green shaded area is the lattice QCD result of 
Refs.~\cite{gergy3, gergy4}. We observe that the red and black curves are in very good agreement for temperatures below $20$ MeV. However, beyond this point there is a significant deviation between the curves, suggesting that the analytical result breaks down rather quickly. While the approximation $(\mu_I-m_{\pi})\ll T$ holds quite well along the black curve displayed in Fig.~\ref{phasediagram} the second approximation $T\ll m_{\pi}$ breaks down somewhere in the lower part of the figure. This elucidates how the discrepancy between the red and black curves can be so significant in the upper region of the figure. 
Finally, we observe that our numerical result is within the uncertainty range of the lattice data for temperatures up to approximately $35$ MeV. This is contrast to the NJL and quark-meson models
that include quark degrees of freedom. The phase boundary predicted by these 
models~\cite{heman2,allofus} is in good 
agreement with lattice results in the entire
temperature range.

\begin{figure}[htb]
\includegraphics[width=0.45\textwidth]{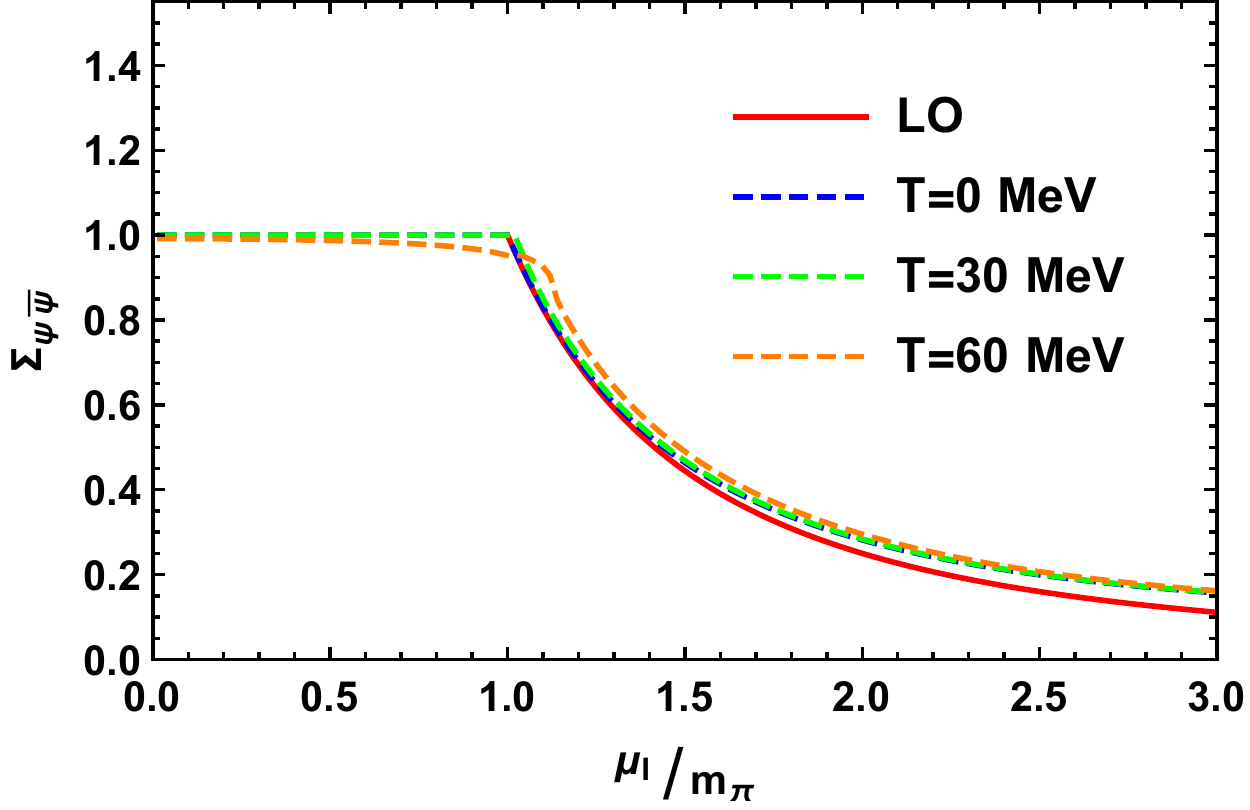}
\caption{Normalized quark condensate deviation, $\Sigma_{\bar{\psi}\psi}$, as a function of the normalized isospin chemical potential at vanishing and finite temperature to one loop. See main text for details.
}
\label{phaselog}
\end{figure}
\begin{figure}[htb]
\includegraphics[width=0.45\textwidth]{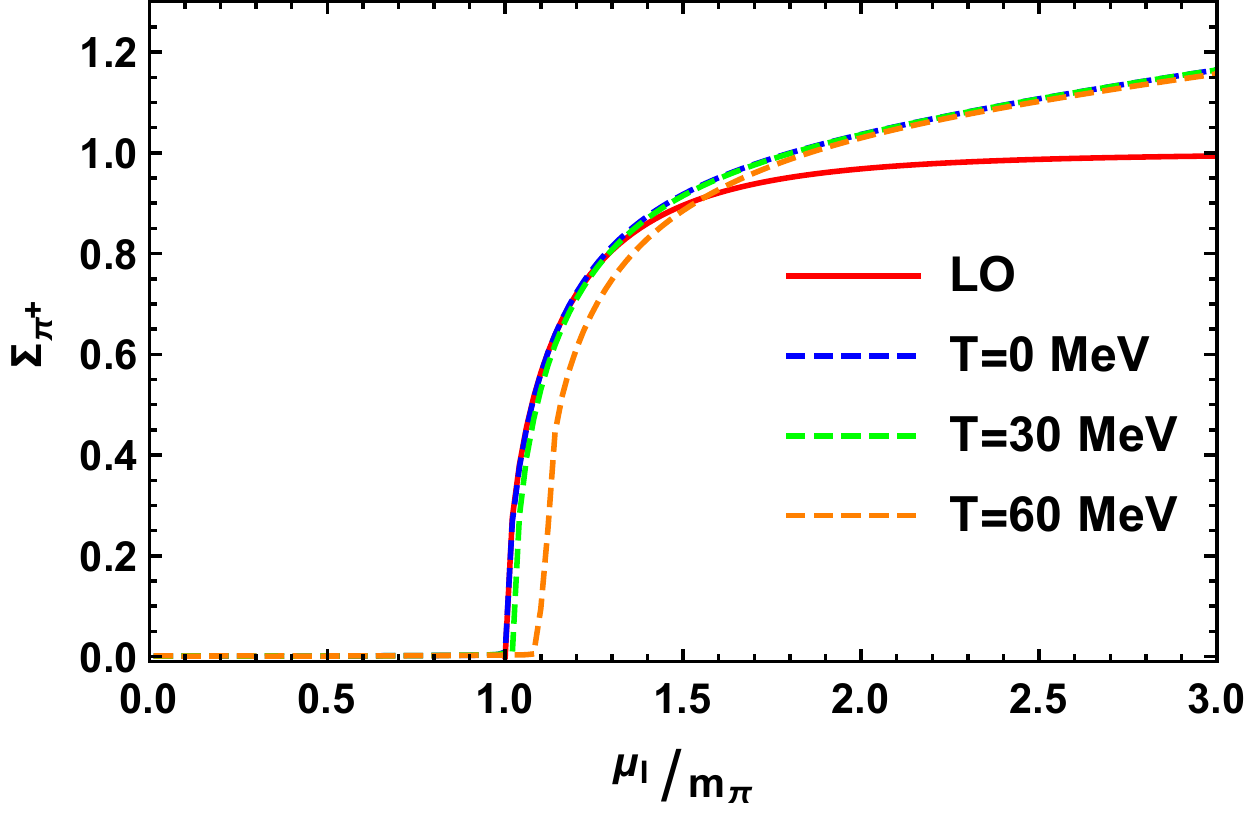}
\caption{Normalized pion condensate deviation, $\Sigma_{\pi^+}$, as a function of the normalized isospin chemical potential at vanishing and finite temperature to one loop. See main text for details.
}
\label{pioncon2}
\end{figure}

Let us next turn to the quark and pion condensates.
In Ref.~\cite{condensates} we compared the $\chi$PT predictions with
those of lattice simulations~\cite{gergy1,gergy2,gergy3} at vanishing temperature.
The comparison was made between condensate deviations, which characterizes the change of condensate relative to the normal vacuum value. These are defined as
\begin{align}
\label{deviation1}
\Sigma_{\bar{\psi}\psi}&=-\frac{2m}{m_{\pi}^{2}f_{\pi}^{2}}\left[\langle\bar{\psi}
    \psi\rangle_{\mu_{I}}-\langle\bar{\psi}\psi\rangle_{0}^{j=0}
\right]+1\;,\\
\Sigma_{\pi^+}&=-\frac{2m}{m_{\pi}^{2}f_{\pi}^{2}}\langle\pi^{+}\rangle_{\mu_{I}}\;.
    \label{deviation2}
\end{align}
At tree level, the definitions ensure that $\Sigma_{\bar{\psi}\psi}^2+\Sigma_{\pi^+}^2=1$, which
simply expresses the rotation (on the chiral circle) of the quark condensate into a pion condensate as $\mu_I$ increases. We note that while the subtraction of $\langle\bar{\psi}\psi\rangle_{0}^{j=0}$
in Eq.~(\ref{deviation1}) eliminates the term involving $\bar{h}_1$, the definition in Eq.~(\ref{deviation2}) is $\bar{h}_1$ dependent. However, the $\bar{h}_1$ dependence in Eq.~(\ref{deviation2}) vanishes automatically when we consider the special case, $j=0$, which is what we do in the following.

In Fig~\ref{phaselog}, we plot $\Sigma_{\bar{\psi}\psi}$ as a function of $\mu_I/m_{\pi}$ for three different temperatures. We also include the leading order result in red valid at $T=0$. The blue line indicates the NLO result at $T=0$, the green line shows the $T=30$ MeV result, and the orange line shows the $T=60$ MeV result. We observe that the magnitude of the chiral condensate decreases as we increase the temperature in the normal phase, as expected. Furthermore, the magnitude of $\Sigma_{\bar{\psi}\psi}$ drops significantly once we enter the pion-condensed phase. We also note that for $T=60$ MeV, the chiral condensate starts decreasing due to thermal effects before the second order phase transition occurs. Furthermore, for a fixed isospin chemical potential the chiral condensate increases with temperature. This occurs due to the delayed onset of pion condensation at higher temperatures. Finally, we note that the chiral condensate deviations at different temperatures approach each other at high isospin densities, where finite-density effects dominate over finite-temperature effects. We observe that including finite-temperature effects for temperatures up to $60$ MeV in $\chi$PT to one loop does not lead to very significant corrections to the zero-temperature result in any part of the phase diagram. 

In Fig.~\ref{pioncon2}, we show $\Sigma_{\pi^+}$ as a function
of $\mu_I/m_{\pi}$ for three different values of the temperature $T$. The red line shows the $T=0$ MeV LO result, the blue line shows the $T=0$ MeV NLO result, the green line shows the $T=30$ MeV NLO
result, and the orange line shows the $T=60$ MeV NLO result.
The green and the blue line are barely distinguishable. The difference between the red line and the orange line in the proximity of the phase transition is quite clear and reflects the pronounced temperature dependence that the phase-transition curve attains at high temperatures in $\chi$PT to one loop, as was clearly shown in Fig.~\ref{phasediagram}. Once again we observe that finite-temperature effects become unimportant at large isospin densities.  
The violation of the relation~(\ref{rotatie})
is clearly seen in Figs.~\ref{phaselog}--~\ref{pioncon2} and 
it has also been observed on the lattice~\cite{gergy3}.
Generally, the agreement between lattice results 
$\chi$PT at $T=0$ improves as one goes from LO to NLO~\cite{condensates}.

\begin{figure}[htb]
\centering\includegraphics[width=0.5\textwidth]{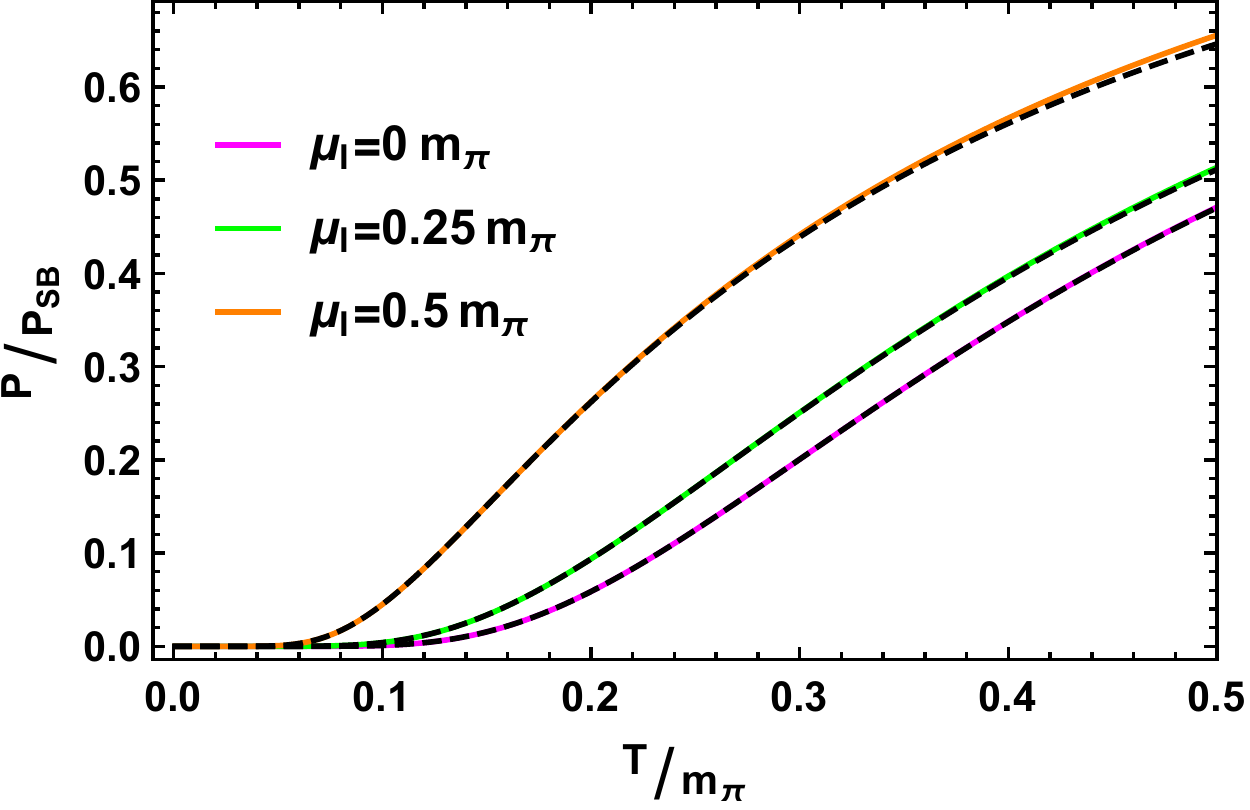}
\caption{Pressure normalized to the Stefan-Boltzmann pressure as a 
function of $T/m_{\pi}$ for three different values of $\mu_I$.
See main text for details.}
\label{trykkj}
\end{figure}

Finally, in Fig.~\ref{trykkj}, we show the pressure divided by the Stefan-Boltzmann
result $P_{\rm{SB}}={T^4\pi^2\over30}$ as a function of $T/m_{\pi}$
for three different values of the isospin chemical potential. From above, $\mu_I=0$, $\mu_I={1\over4}m_{\pi}$, and $\mu_I={1\over2}m_{\pi}$. The pressure at $T=0$ has been subtracted. 
The solid lines are the one-loop results, while the black dashed lines display the two-loop result. The low-energy expansion seems to be converging very well. The good convergence properties of the low-energy expansion were
also found for the quark condensate up to three loops at $\mu_I=0$~\cite{gerber}.

While it is clear that $\chi$PT is breaking down at
some temperature below the deconfinement temperature simply
because it only incorporates mesonic degrees of freedom, rigorous
statements about the validity of our calculations seems difficult since there are several
scales involved, namely $m_{\pi}$, $f_{\pi}$, $T$, and $\mu_I$. 
Considering the green band in Fig.~\ref{phasediagram},
comparing $\chi$PT and lattice QCD, 
suggests that $\chi$PT ceases to be valid at approximately 40 MeV near the phase transition to the pion-condensed phase. A more conservative estimate would be half of this, 20 MeV, where our result for the phase-transition curve is in very good agreement with the analytical approximation of Ref.~\cite{split2}. 

\section*{Acknowledgements}
 The authors would like to thank B. Brandt, G. Endr\H{o}di and S. Schmalzbauer
for useful discussions as well as for providing the
data points of Refs. \cite{gergy3, gergy4}. 

\appendix
\section{Integrals and Sum-integrals}
The sum-integral is defined as
\bqa
\sumint_P&=&T\sum_{n}\int_p=T\sum_n
\left({e^{\gamma_E}\Lambda^2\over4\pi}\right)^{\epsilon}
\int{d^dp\over(2\pi)^d}\;,
\eqa
where the sum is over the Matsubara frequencies $P_0=2\pi n T$.
The integral over three-momenta is regularized using dimensional
regularization with $d=3-2\epsilon$, where $\Lambda$ is
a renormalization scale associated with 
the $\overline{\rm MS}$ scheme.
In dimensional regularization, we find
\bqa
\int_p\sqrt{p^2+m^2}&=&-
{m^4\over2(4\pi)^2}\left({\Lambda^2\over m^2}\right)^{\epsilon}
\left[{1\over\epsilon}+{3\over2}+{\cal O}(\epsilon)\right]\;.
\eqa
\begin{widetext}
The sum-integrals we need are
\bqa
\sumint_P\log\left[(P_0+i\mu_I)^2+p^2+m^2\right]&=&
\int_p\sqrt{p^2+m^2}+
T\int_p\left\{\log\left[1-e^{-\beta(E_p-\mu_I)}\right]
  +\log\left[1-e^{-\beta(E_p+\mu_I)}\right]
\right\},
\label{sumint22}
\\
\sumint_P{1\over(P_0+i\mu_I)^2+p^2+m^2}&=&
-{m^2\over(4\pi)^2}\left({\Lambda^2\over m^2}\right)^{\epsilon}
\left[{1\over\epsilon}+1+{\pi^2+12\over12}\epsilon+{\cal O}(\epsilon^2)\right]+
{1\over2}  \int_p\left[{n_B(E_p-\mu_I)\over E_p}+{n_B(E_p+\mu_I)\over E_p}\right]\;,
 \\ 
\sumint_P{(P_0+i\mu_I)
  \over[(P_0+i\mu_I)^2+p^2+m^2]}&=&{i\over2}
  \int_p\big[n_B(E_p-\mu_I)-n_B(E_p+\mu_I)\big]\;,
  \label{sumint33}
  \eqa
\end{widetext}
where 
$n_B(x)={1\over e^{\beta x}-1}$ is the Bose-Einstein distribution
function
and $E_p=\sqrt{p^2+m^2}$.
The sum over Matsubara frequencies in the
sum-integrals
can be performed using contour integration with standard techniques, see
e.g. Ref.~\cite{kapusta} for details.
We note that the sum-integral in Eq.~(\ref{sumint33}) is finite, and 
vanishes in the limit $\mu_I\rightarrow0$
since it is odd in the variable $P_0$.

\end{document}